# The phenomenon of growing surface interference explains the rosette pattern of jaguar

Michaël Dougoud[1], Christian Mazza[1], Beat Schwaller[2], Laszlo Pecze[2]

[1] Department of Mathematics, University of Fribourg, Chemin du Musée 23, CH-1700 Fribourg, Switzerland

[2] Anatomy, Department of Medicine, University of Fribourg, Route Albert-Gockel 1, CH-1700 Fribourg, Switzerland

To whom correspondence should be addressed: Laszlo Pecze, Anatomy, Department of Medicine, University of Fribourg, Route Albert-Gockel 1, CH-1700 Fribourg, Switzerland.
Tel. ++41 26 300 85 11  Fax: ++41 26 300 97 33, E-mail: laszlo.pecze@unifr.ch

# ABSTRACT

One possible mechanism to explain how animals got their coat patterns was proposed by Alan Turing. He assumed that two kinds of morphogens diffuse on a surface and interact with each other, generating a reaction-diffusion mechanism. We developed a new framework for pattern generation incorporating a non-diffusing transcription factor in the system. The diffusion factors (one inhibitor and one activator) acting on cell surface receptors modulate the activity of a transcription factor. The difference in the local concentration of diffusion factors is translated into the degree of activation of transcription factors. The speed of this process determines then pattern formation velocity, i.e. the elapsed time from an initial noisy situation to a final developed pattern. If the pattern formation velocity slows down compared to the growth of the surface, the phenomenon of "growing surface interference'' occurs. We find that this phenomenon might explain the rosette pattern observed on different types of felids and the pale stripes found between the regular black stripes of zebras. We also investigate the dynamics between pattern formation velocity and growth and to what extent a pattern may freeze on growing domains.

**Introduction**

A rosette is a characteristic marking found on the surface of some animals, particularly on the fur coat of feline predators. Rosettes are likely used for camouflage purposes rather than for communication or other physiological reasons (Allen et al., 2011). They help predators to deceive the prey by simulating the different shifting of shadows, helping them to remain visually hidden. Rosettes exist with or without central spots; however the central color tone is darker than the background color. Felids displaying rosettes include jaguar (*Panthera onca*), leopard (*Panthera pardus pardus*), snow leopard (*Panthera uncia*), ocelot (*Leopardus pardalis*), leopard cat (*Prionailurus bengalensis*) and Bengal cat (*Prionailurus bengalensis X Felis catus*). Generally, newborn animals do not have rosettes, but instead regular black dots; the pigmentation pattern changes as they grow. The spots turn initially to rings and then develop to rosettes.

Coloration on the surface of animals (skin, fur) is determined by the distribution of specialized pigment cells called melanocytes (Simon and Peles, 2010). The reaction-diffusion model proposed by Alan Turing (Turing, 1952) explains how spatial patterns may develop autonomously. Stimulation of melanocytes responsible for pigment synthesis to a given site is under the control of a diffusible compound called the activator, which also stimulates its own production through a positive feedback loop. In order to form a pattern, an additional mechanism (inhibitor) is needed for suppressing the production of the activator in the neighborhood of the autocatalytic center. Thus, the pattern is formed as a result of the antagonistic interaction between short-range activators and wide-range inhibitors. While this theory has been widely explored since Turing's seminal work (Turing, 1952), giving rise to different modeling approaches and analytical results on pattern formation (Maini et al., 2006; Miyazawa et al., 2010; Murray, 1982), only recent advances have permitted to identify the exact nature of chemical compounds acting as pairs and satisfying the model's requirements. The interaction between two morphogens named fibroblast growth factor (FGF) and sonic hedgehog dictate the ridge patterns in the mouth of mice, as predicted by Turing's models (Economou et al., 2012). Different variants of reaction-diffusion models such as Alan Turing's linear model (Economou et al., 2012), Gierer–Meinhardt (Gierer and Meinhardt, 1972), Gray–Scott (Gray and Scott, 1984) and BVAM model (Barrio et al., 1999) have been developed for simulation of regularly spaced dots, labyrinths and stripes. Previous computer simulations had evaluated the effect of a growing domain in one-dimensional space using different methods, such as insertion (Arcuri and Murray, 1986; Kondo and Asal, 1995) or continuous domain growth, either being uniform (Crampin and Maini, 2001 ) or non-uniform (Crampin et al., 2002; Neville et al., 2006). Moreover, studies of growing domains in two or more dimensions have been reported. In the paper of Madzvamuse (Madzvamuse, 2006) isotropic and uniform growth was investigated on two-dimensional domains and analytical insights were provided (Madzvamuse et al., 2010). Results on growing domains with curvature, e.g. spheres have also been investigated (Gjorgjieva and Jacobsen, 2007; Iber and Menshykau, 2013). Very recently the dynamics of growth (slow versus fast growth) has been explored; this work highlights the difficulties of predictions in the case of fast growth (Klika and Gaffney, 2017). Besides these mean-field continuous systems, stochastic systems descriptions have also been investigated (Woolley et al., 2011a, b). Although the abovementioned models can be also used to examine the effect of a growing surface, we propose here a new model. It postulates that the effect of diffusion factors has to be transmitted via cellular signaling transduction pathways resulting in the activation of a transcription factor (TF) (Mallarino et al., 2016). The diffusion factors bind to their corresponding receptors usually localized on the surface of a cell. This induces a signaling cascade that leads to the shuttling of the TF between the cytosol and the nucleus (Cai et al., 2008; Nakayama et al., 2008; Nelson et al., 2004). If the TF is retained in the nucleus for a sufficiently long period of time, then it stimulates the translation of the "color genes" in the skin, whose product(s) act(s) on pigment-producing melanocytes (Abdel-Malek and Swope, 2011; Horikawa et al., 1995) and the production of diffusion factors. Assuming that cellular signaling transduction pathway influences how fast the difference in the local concentration of diffusion factors is translated into the degree of the activation of transcription factor, the third non-diffusing compound allows for modifying the velocity of the reaction-diffusion system. If the velocity of the pattern formation is slow compared to the growth rate of the embryo or a young animal, then the phenomenon of "growing surface interference" occurs. Note that a third non-moving substance has also been considered in some mathematical work (Klika et al., 2012; Marcon et al., 2016; Raspopovic et al., 2014), investigating the presence of cell-autonomous factors. But the effect of growth was not examined in these studies.

We show that this phenomenon is likely to be involved in the pattern formation of animal coats such as the rosette patterns on different types of cats, e.g. Bengal cat (Fig. 1B) and rays e.g. Rosette river stingray (*Potamotrygon schroederi*) (Fig. 1C). This effect also accounts for the pale stripes between the regular black stripes of zebras (*Equus zebra*) (Fig. 1A). Our analysis relies on decreasing reaction velocities of pattern formation with time, which enables to "freeze" the patterning.

## Model

### General considerations

Activating (A) and inhibiting (I) diffusion factors acting as morphogens may consist of intracellular molecules such as mRNAs or miRNAs that set up a concentration gradient by diffusion in a syncytium or in cells connected with gap junctions. But more commonly, morphogens comprise secreted proteins forming an extracellular gradient across a field of cells (Christian, 2012). Diffusion factors binding to the corresponding receptor(s) and eliciting a signaling cascade influence the degree of the activation level of transcription factor (TF). The activated TF translocates to the nucleus and I) stimulates the production of a paracrine factor ("color factor") that influence the pigment synthesis in melanocytes and II) leads to the synthesis of the activator and the inhibitor.

### Mathematical modeling without growth

Let $A(t, x)$ denote the concentration of activator, $I(t, x)$ the inhibitor concentration and $S(t, x)$, the degree of activation of transcription factors (TF) present in the nucleus at time $t$ and place $x$. This can also be seen as a proxy for the expression level of the "color factor" within the system. Considered here will be two-dimensional (i.e., $x = (x_1, x_2)$) domains. We rely on reaction-diffusion equations for the activators and inhibitors of the following form,

$$\frac{\partial A(t,x)}{\partial t} = r_{ab}(t)\, b_a\, S(t,x) - d_a\, A(t,x) + D_a\, \nabla^2 A(t,x) + A(t,x) * \xi_a \tag{1}$$

$$\frac{\partial I(t,x)}{\partial t} = r_{ab}(t)\, b_i\, S(t,x) - d_i\, I(t,x) + D_i\, \nabla^2 I(t,x) + I(t,x) * \xi_i, \tag{2}$$

where $b_a$ is the production rate, $d_a$ the degradation rate, and $D_a$ the diffusion coefficient of the activator, while $b_i$, $d_i$, and $D_i$ are the production rate, degradation rate and diffusion coefficient of the inhibitor. The random variables $\xi_a$ and $\xi_a$ are distributed normally $\xi_a \sim \mathcal{N}(0, \sigma_a^2)$ and $\xi_i \sim \mathcal{N}(0, \sigma_i^2)$, in line with the study of Zheng et al., (Zheng et al., 2017). If not noted otherwise, $\sigma_a^2 = \sigma_i^2 = 0$, i.e. there is no stochastic process in the system. All these parameters are non-negative constants. The parameter $r_{ab}$ allows the change in the production rates of the inhibitor and activator in the same manner. If not specified, we choose it to be $r_{ab}(t) \equiv 1$. We assume here that TF promote the production of both diffusion factors. The dynamics of transcription factors is implemented by the following differential equation,

$$\frac{dS(t,x)}{dt} = r_v(t) \cdot \left( b_s\, \frac{(A/I)^2}{K+(A/I)^2} - d_s\, S(t,x) + r_p \right), \tag{3}$$

where $b_s$ is the rate of transcription factors conveyed to the nucleus, $d_s$ the rate of transcription factors removed from nucleus, $r_p$ is a small random rate of transcription factors always transported into the nuclues. Thus, the value $S$ can be regarded as the degree of activation of TF, if $A \gg I$, $S \approx (b_s + r_p)/d_s$ and if $I \gg A$, $S \approx r_p/d_s$. The function $r_v(t)$ is the reaction velocity. It represents the effectiveness of the system to translate local differences in the diffusion factors into the degree of activation of TF. If not specified, we will consider it to be such that $r_v(t) \equiv 1$. The function $r_v(t)$ is the reaction velocity, assumed to be decreasing with time with maximal value 1 at time $t = 0$. It represents the time required for the system to translate the local differences in the diffusion factors into the degree of activation of TF. In our simulations we relied on a Hill-type function,

$$r_v(t) = \begin{cases} 1, & t < t_0 \\ \frac{k_v^{\eta}}{k_v^{\eta}+(t-t_0)^{\eta}}, & t \geq t_0 \end{cases} \tag{4}$$

where $t_0$ is the time at which the reaction velocity begins to decrease, $\eta \geq 1$ the Hill coefficient, and $k_v > 0$ the half-saturation constant. A schematic representation of the recent and the new modeling framework is presented in Fig. 2. Linear stability analysis for this system has been performed in another study (M. Dougoud et al., submitted) showing that diffusion-driven instability occurs only when $D_a < D_i$. We take the initial conditions ($t = 0$) for $A$, $I$, and $S$ to be random, uniformly distributed on the intervals $A_{ini}$, $I_{ini}$, and $S_{ini}$ respectively. We also test an initial tendency in the amount of transcription factors (linear vertical gradient, from some parameter $S_{max} \geq 0$ to $S_{min} \geq 0$).

**Assumptions on growth and Lagrangian framework**

To incorporate growth in equations (1-3), we consider the flow velocity $V(t, x)$ at position $x$ and time $t$ generated by domain growth and follow the works of Maini and collaborators (Crampin et al., 2002; Madzvamuse et al., 2010; Madzvamuse and Maini, 2007). This flow introduces a growth term in every equation (1-3) of the form $\nabla(VW)$, where $W \in \{A, I, S\}$ depending on the equation. In the following we investigate growth on two-dimensional planar domains. Equations (1-3) become

$$\frac{\partial A(t,x)}{\partial t} + \nabla(VA) = r_{ab}(t)\, b_a\, S(t,x) - d_a\, A(t,x) + D_a\, \nabla^2 A(t,x) \tag{5}$$

$$\frac{\partial I(t,x)}{\partial t} + \nabla(VI) = r_{ab}(t)\, b_i\, S(t,x) - d_i\, I(t,x) + D_i\, \nabla^2 I(t,x), \tag{6}$$

$$\frac{dS(t,x)}{dt} + \nabla(VS) = r_v(t) \cdot \left( b_s\, \frac{(A/I)^2}{K + (A/I)^2} - d_s\, S(t,x) + r_p \right). \tag{7}$$

Note that with growth, position $x$ depends on $t$. We use therefore Lagrangian coordinates to map the deforming domain onto a fix domain, see Fig. 2C. This deformation is assumed to be continuous, uniform, and isotropic. With these assumptions, in two dimensions, we can map the position $x(t)$ on a fixed position $\xi = (\xi_1, \xi_2)$ (lying on the fixed domain) such that

$$x(t) = \rho(t)\, \xi, \tag{8}$$

where $\rho(t)$ is a growth factor describing the domain's deformation with time (Iber and Menshykau, 2013; Madzvamuse and Maini, 2007). We map then $A$, $I$, and $S$ in this particular framework. Using derivations rules (Madzvamuse and Maini, 2007), the following differential equations are obtained and represent the model in its most general form,

$$\frac{\partial A(t,\xi)}{\partial t} = r_{ab}(t)\, b_a\, S(t,\xi) - d_a\, A(t,\xi) + \frac{D_a}{(\rho(t))^2}\, \nabla^2 A(t,\xi) - 2\frac{\dot{\rho}(t)}{\rho(t)} A(t,\xi) \tag{9}$$

$$\frac{\partial I(t,\xi)}{\partial t} = r_{ab}(t)\, b_i\, S(t,\xi) - d_i\, I(t,\xi) + \frac{D_i}{(\rho(t))^2}\, \nabla^2 I(t,\xi) - 2\frac{\dot{\rho}(t)}{\rho(t)} I(t,\xi) \tag{10}$$

$$\frac{dS(t,\xi)}{dt} = r_v(t) \cdot \left( b_s\, \frac{\left(\frac{A}{I}\right)^2}{K + \left(\frac{A}{I}\right)^2} - d_s\, S(t,\xi) + r_p \right) - 2\frac{\dot{\rho}(t)}{\rho(t)} S(t,\xi). \tag{11}$$

Note that the last term of each equation accounts for dilution in the system due to growth. Relying on the paper Madzvamuse (Madzvamuse et al., 2010) we mostly use two types of deformation: the linear and the logistic types. The linear type is given by

$$\rho_{lin}(t) = r_g\, (t - t_s) + 1 \tag{12}$$

where $t_s \geq 0$ is the time at which growth starts and $r_g \geq 0$ the growth rate. The logistic type has the following particular form,

$$\rho_{log}(t) = \frac{1 + \kappa\, e^{r_g\, (t - t_i)}}{1 + e^{r_g\, (t - t_i)}}, \tag{13}$$

with $t_i \geq 0$ the time at which $\rho_{log}$ has an inflexion point, $\kappa \geq 1$ the asymptotic growth factor and $r_g \geq 0$ the growth rate.

**Simulations**

All the simulations were performed with MATLAB R2012b (MathWorks Inc., MA). The MATLAB code for Turing pattern generation as presented by Jean Tyson Schneider (Schneider, 2012) was used as an initial framework for our program. We used finite differences with a time step set to $\Delta t = 0.01$. The reference fixed spatial domain is a square of size $M$ and has been discretized such that $\Delta x = \Delta y = 1$, were the parameter usually taken in our simulations is $M = 100$. Equations (9-11) are treated with no-flux boundary conditions.

In our two-dimensional domain model, the velocity of the pattern formation is followed by changes of the standard deviations, $\sigma(t)$, of the $S(t, \xi)$ values,

$$\sigma(t) = \sqrt{\frac{1}{\widehat{M}^2} \sum_{i=1}^{\widehat{M}^2} (S(t, \xi_i) - \bar{S}(t))^2}, \tag{14}$$

where $\widehat{M}$ is the number of one-dimensional spatial steps in our simulations and $\bar{S}(t)$ is the empirical mean of the values of $S(t, \xi)$ on the domain at time $t$. In our simulations, we have then tracked the evolution of $\sigma(t)$, which is a good indicator of the convergence state of the process. In the end, higher standard deviations will be related to more developed patterns with clearly separated colors. Its derivative represents thus the speed at which a pattern develops and the analysis of the maximum of $\sigma'(t)$ will permit to highlight prominent parameters involved in pattern freezing and development.

**Results**

1. **Analysis of reaction velocity without growth**

Our model is capable of producing labyrinth patterns in a certain range of parameters. The velocity of the pattern formation is followed by changes in the standard deviations of $S(t, x)$. Developed patterns produce the highest asymptotic $\sigma(t)$ values, if the initial noise is reduced. From the noisy initial conditions, a pattern is formed following a curve of the Hill form of standard deviation values with an inflection point around time $t = 250$, if $r_v = r_{ab} = 1$, i.e. reaction velocity and the simultaneous rate of diffusion factors' production do not decrease with time. When the parameter of reaction velocity is decreased, as e.g. with $r_v \equiv 0.5$, then the inflection point of the Hill equation is increased to a value of around $t = 360$ iteration time. Compared to Fig. 3A, one observes greyer regions in Fig. 3C. However at the end stage, both simulations result in the same pattern type (compare Fig. 3B to Fig. 3D). This indicates that with the parameter $r_v$, seen as a velocity of the intracellular signaling pathways leading to the activation of TFs, one can regulate the speed of the pattern formation. If this value is close to zero the pattern is getting frozen and will not change any longer. More generally, larger values of $r_v$ lead to steeper $\sigma(t)$, i.e. at some time point the speed of pattern development will be faster when $r_v$ is large, see Fig. 3H. It is worth noting that the existing reaction-diffusion models using only two differential equations are capable to produce this phenomenon only when the production rates and the diffusion coefficients for activator and inhibitors are changing in the same synchronized manner. From a biological viewpoint this is a very unlikely situation. Otherwise any modulation of one of the parameters of existing models will produce another type of pattern(Miyazawa et al., 2010). Also, if we perform an adiabatic reduction on our model, reducing it into two differential equations, we lose the $r_v$ parameter and the speed of the pattern formation cannot be easily regulated (Supplementary Information).

When the production rates of $A$ and $I$ are simultaneously decreased with $r_{ab} = 0.5$, no changes in the activation levels of TFs, in the speed of the pattern formation or in the form of the pattern may be observed (Fig. 3A,B *vs.* Fig. 3E,F).

The changes of $\sigma(t)$ during the process are shown in Fig. 3G. An analysis of the maximum of $\sigma'(t)$ (the steepness of the slope) in response to different $r_v$ values shows a Hill-type curve. This indicates that the pattern formation velocity asymptotically approaches a theoretical maximum (Fig. 3F).

2. **Analysis of the effect of noise on the reaction velocity without growth**

In our model, we distinguish initial noise (uniform distribution of $A_{ini}$, $I_{ini}$ and $S_{ini}$) in the concentrations of diffusion factors and TFs, which leads essentially to different initial conditions and a white noise involved in the production of diffusion factors $A$ and $I$. If not noted otherwise, $\sigma_a^2 = \sigma_i^2 = 0$, i.e. there is no stochastic process in the system. Increasing the initial noise in the concentrations of TFs does not change the curve of $\sigma(t)$ values ($\sigma_a^2 = \sigma_i^2 = 0$). However, an increase in $S_{ini}$ promotes the pattern formation ($\sigma_a^2 = \sigma_i^2 = 0$). Adding white noise ($\sigma_a^2 = \sigma_i^2 > 0$) slightly slows down the speed of the process; these results are illustrated in Fig. 4B. In this case, to obtain the numerical differentiation of the noisy data, least-square polynomial approximations were performed (Knowles and Renka, 2014). We used local regression using weighted linear least squares and a 2$^{nd}$ degree polynomial model (LOESS) with 10% span (percentage of the total number of data points). From the smoothened curve, the maximum value of the first difference was calculated.

3. **Analysis of growing surface interference**

In our model we investigated the effect of the growth rate $r_g$ and the reaction velocity $r_v$ on the development of patterns shown in Fig. 5 with illustrations shown in Fig. 6. As shown in previous systems producing the prototypical Turing patterns(Miyazawa et al., 2010), changing one parameter produces regularly spaced black or white dots in a background of the opposite color and labyrinth patterns in between. For analysis, we selected a parameter set producing regularly spaced black dots on a white background and we chose a linear growth in the model. Linear growth represents well the observed growth rate during the middle and late embryonic stages (Mu et al., 2008) or in juvenile ages (Lamonica et al., 2007). As shown in Fig. 5A, an increased growth rate negatively influences the steepness of the curve $\sigma(t)$, i.e. interferes with pattern formation velocity. Moreover the pattern itself is also altered; the pattern shows more and more curved lines instead of regular dots. The selected growth rates ($r_g = 0$ , $r_g = 0.002$ and $r_g = 0.004$ are presented in Fig. 5B and the resulted patterns are shown in Fig. 5C, D and E, respectively. Fig. 5F shows that a larger reaction velocity $r_v$ and a smaller growth rate $r_g$ lead to local rapid increases of $\sigma(t)$ and mutually influence each other.

Varying these two factors we may obtain very similar patterns as observed on the adult skin of animals. Since the growth rate slows down and even stops in early adulthood for many animals, we used logistic growth rates for simulations. The involvement of concentration gradients has already been proposed before to play a role in stripe formation(Hiscock and Megason, 2015). In our model, if one uses a linear initial gradient for the activation level of TFs, the system forms regular stripes and the slope of the gradient determines two poles. The onset of stripe generation is localized at the negative pole. Later on stripes develop at the positive pole, until the middle of the surface begins to be inhabited by further stripes. The size of the surface determines the number of stripes. Only few stripes may be formed very fast, near the poles. In order to generate more stripes, new stripes may appear from the division of older ones in a period of growth. White stripes are inserted on the skin of the marine angelfish *Pomacanthus* (Kondo and Asal, 1995), while black stripes are inserted in the case of the zebra. During growth in our model (Figs. 6A-D) a transition state occurs generating a narrower white stripe periodically after each second black stripe (Fig. 6B). If the reaction velocity decreases during this process, the transition state remains stable and it will not be sensitive to additional growth of the domain (Figs. 6E-H). Note that in this case, the $\sigma(t)$ values are reduced when reaction velocity slows down and stabilize when the domains stop growing (Fig. 6H, blue line).

In a certain parameter range our model produces regularly spaced dots. If the domain growth is relatively slow (Figs. 6I-L) compared to the reaction velocity, new dots are generated by division of the existing ones (Fig. 6K). However, if the domain growth is fast compared to reaction velocity (Figs. 6M-P), the dots do not have enough time to divide regularly. They transform under this mechanical (surface increasing) process into doughnut shapes, which afterwards convert to rosette patterns. This process predicted by our model is highly similar to the age-dependent changes of the patterns observed on the fur of a jaguar; the same animal was photographed at different ages (Fig. 1D-F). Note that in some cases, with sufficiently fast growth, black dots also appear in the middle of the rosettes (see Fig. 5E).

**Discussion**

The coloration of animal skin is due to melanin pigments that are produced by melanocytes. Melanocytes are located in the *stratum basale* layer of the skin's epidermis and in hair follicles; melanocytes secrete mature melanosomes to surrounding keratinocytes(Lin and Fisher, 2007). The localized changes in the homogeneous distribution of different types of melanocytes (in non-mammalian vertebrates) or in the pigment synthesis pathway (in mammals) result in different patterns (Mills and Patterson, 2009). In felids pigment-type switching controlled by *Asip* and *Edn3*, both factors acting in a paracrine way, is the major determinant of color patterns (Kaelin et al., 2012). These factors originate from the dermal papilla of hair follicles and influence the pigment synthesis pathway *leading to the production of either eumelanin (brown-black)* or pheomelanin (yellow-red) (Mills and Patterson, 2009). The production of *Asip* is promoted by the morphogen BMP-4 (Abdel-Malek and Swope,

2011). BMP-4 is assumed to work as an inhibitory molecule in several Turing-type developmental processes (Miura, 2007). In African striped mice (*Rhabdomis pumilio*) and Eastern chipmunks (*Tamias striatus*) the periodic dorsal stripes are the result of differences in melanocyte differentiation and the transcription factor ALX3 has been found to be a regulator of this process (Mallarino et al., 2016).

Rosette-pattern formation on felines has been explained before based on a mathematical model (Liu et al., 2006). In this model, rosette-like patterns are generated by tuning various parameters of the model during its temporal evolution. Nevertheless, there are several points during pattern formation *in vivo* (Fig. 1D-F) that might require a modification of the model by Liu. Namely: 1) On the jaguar coat, the dots transform into rosettes on the animal's body, but remain dots on the face of the animals. As growth proceeds, the head occupies a decreasing proportion of body size (surface) both in humans and animals (Alley, 1983). Thus, the growing surface interference is more pronounced on the body and accounts for its effect on patterning. 2) In our framework, only domain size changes permit the appearance of different patterns (see Fig. 6A-P), in line with the observed growth of animal coat surfaces. It thus avoids tuning of parameters during the growth process, likely resulting in a more robust process insensitive to perturbations. It further allows optionally to controlling parameters by making use of a decreasing reaction velocity with time under certain circumstances.

In Turing systems the number of peaks or lines increases with larger domain size. Turing systems with two components have been analyzed with fixed reaction velocity, but during the process of pattern development the domain size may change. This effect was examined in several studies. White stripes are inserted on the skin of the marine angelfish *Pomacanthus* (Kondo and Asal, 1995) in the period of its juvenile growth, but the distance between the stripes does change. Thus, the number of stripes increases proportionally to body size: if the field length reaches about twice the original length, new stripes appear in the middle of the original stripes. This was simulated by a reaction-diffusion wave on a growing array of cells. One of the five cells is forced to duplicate periodically (Kondo and Asal, 1995). Analyzing the domain growth in a one-dimensional space revealed that robust doubling of peaks (when the new peak is always developing in the middle of the existing peak retaining their original mode) is only realized within a specific range of growth rates. Thus, a too slow domain growth results in asymmetric peak splitting. However, if growth is too fast, mode doubling failure occurs, because the system does not remain for sufficiently long time in a given domain range to allow for the establishment of the intended pattern (Barrass et al., 2006; Maini et al., 2012). The apparition of new stripes due to domain growth certainly plays a role in the development of zebra stripes as some zebras have frozen transitional patterns; periodically repeating narrow and wide black stripes.

Our model embracing the reaction velocity, $r_v(t)$ parameter, allows to modifying the pattern formation velocity. Based on the $\sigma(t)$ curves (see Fig. 3G) we can distinguish several phases of the pattern formation in analogy to the growth of bacteria or other microorganisms in batch culture: 1) An initial noisy situation is followed by a short lag phase, when the $\sigma(t)$ values do not change significantly and cells adapt themselves to pattern formation, 2) The log phase is a period characterized by fast increases in the $\sigma(t)$ values and quickly forming patterns. We may call these patterns as transient patterns and 3) The stationary phase, when the $\sigma(t)$ parameter does not change significantly. We may call these patterns: fully developed patterns. For example regularly-spaced dots are a type of fully developed pattern. Our model shows that increased growth rate and lower reaction velocity $r_v(t)$ values can inhibit the appearance of a fully developed pattern and transient patterns may be present on an animal's skin. The jaguars' rosetta is a good example of what we call a transient pattern.

In conclusion, we have developed a rather simple model, whose main relevance concerns the effect of the temporal growth of domains and the pattern formation velocity. This feature is able to shed new lights on different common patterning observed on felids. Beyond parameter fine-tuning we have provided a framework to explain that such phenomena may be the consequence of a mechanical process occurring during the growth of individuals embracing so apparently distinct process as rosette formation and sequentially repeating fainted stripes on zebras.

**Acknowledgements**
Photos courtesy of Dr. Guy Castley, Mark Henry Sabaj Pérez, ZOO Gyor and Helmi Flick. The authors would like to thank Dr. Girard Franck, University of Fribourg for carefully reading our manuscript and for giving his constructive comments.

**Conflict of Interest**
The authors declare no conflict of interest

**Funding**
There was no external funding supporting this study.

**Authors contribution**
LP conceived the study MD CM LP designed the model MD LP performed simulations and data analysis MD BS LP wrote manuscript.

Figure legends

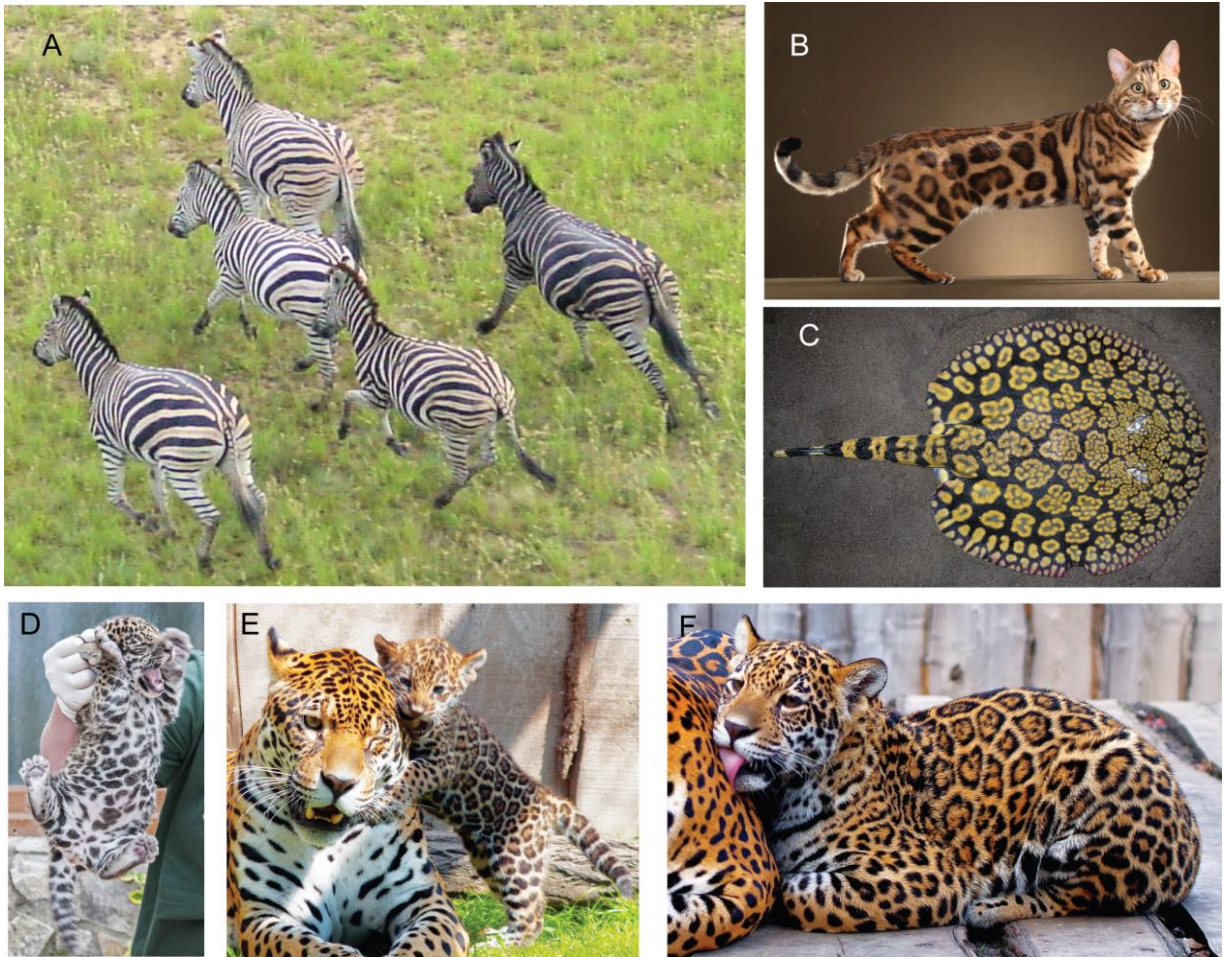

**Fig. 1 Animal skin patterns. A**) A group of zebras (*Equus zebra*). Each zebra has its own irregular patterning. The widths of the black lines are different in each consecutive stripe. Sometimes the color of the thin black lines is weaker than that of the wide black lines. The zebra on the left is even more special, having very narrow white lines. Photographed by Dr. Guy Castley. **B**) A Bengal cat (*Prionailurus bengalensis X Felis catus*) showing some long drawn out rosettes. Photographed by Helmi Flick. **C**) Rosette river stingray (*Potamotrygon schroederi*). Photographed by Mark Henry Sabaj Pérez. **D-F**) Jaguar (*Panthera onca*). Photos courtesy of ZOO Gyor. **D**) 2-weeks old, **E**) 3-months old (right), and **F**) 8-months old

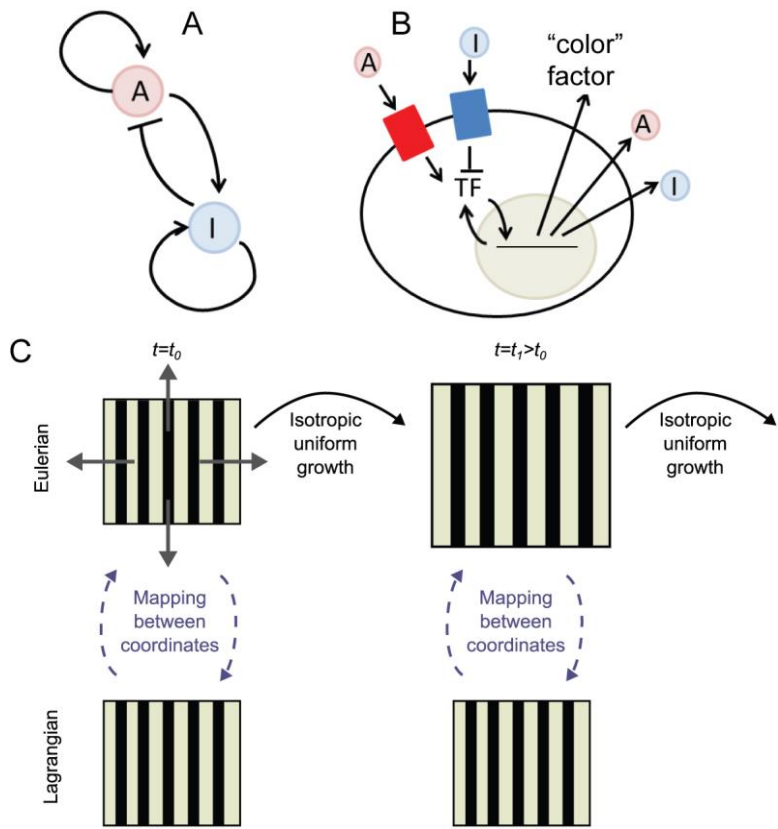

**Fig. 2 Schematic representation of the models: A)** The original model of Turing(Turing, 1952) considers two diffusing factors, an inhibitor (I) and an activator (A), with a requirement of long-range inhibitory effects. **B)** The new model assumes that cells must translate the diffusing-factor encoded information into a biological signal. Diffusion factors binding to the corresponding receptors (red and blue rectangles) activating or inhibiting the transcription factor (TF). The activated TF translocates to the nucleus and I) stimulates the production of a skin-color influencing factor and II) leads to the synthesis of the activator and the inhibitor. **C)** Illustration of the difference between an Eulerian and a Lagrangian framework with isotropic uniform growth.

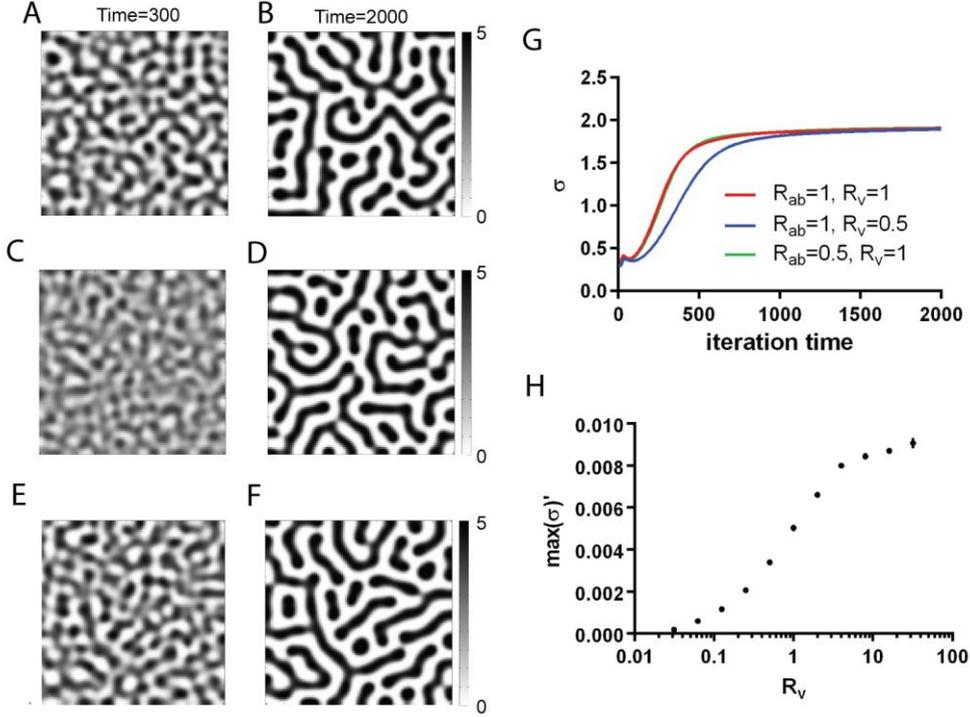

**Fig.3. Illustration of the effect of reaction velocities on pattern development.** The velocity of the pattern formation is followed by changes of the standard deviations of the $S(t,x)$ values. $\sigma(t,x)$, The degree of activation of transcription factor $S(t,x)$ can also be seen as a proxy for the expression level of the "color factor" within the system. **A-B)** $r_v(t) = r_{ab}(t) \equiv 1$, **C-D)** $r_v(t) \equiv 0.5$, $r_{ab}(t) \equiv 1$, and **E-F)** $r_v(t) \equiv 1$, $r_{ab}(t) \equiv 0.5$ at running times $t = 300$ (A,C,E) and $t = 2000$ (B,D,F). **G)** The standard deviation, $\sigma(t)$, of the three simulated processes (A-B) red, (C-D) blue and (E-F) green traces as a function of time. The three $\sigma(t)$ curves converge to the same value as time increases. The second process is slower. All parameters are given in Table 1. $A(0,x)$, $I(0,x)$ and $S(0,x)$ are uniformly sampled on $[0.5, 0.6]$, $[0.5, 0.6]$, and $[0.5, 1.5]$ respectively. The size of the domain is fixed. The pattern type is not affected by a simultaneous change of the production rates of diffusion factors. H) The steepness of $\sigma(t)$ curves, characterized by the maximum of $\sigma'(t)$ values, in response to $r_v$ values, asymptotically approaches a theoretical maximum, the maximum of the pattern formation velocity. Data represent mean±standard deviation out of 3 independent simulation runs.

|   | **Reaction** | **Diffusion** | **Other parameters** |
|---|---|---|---|
| S | $b_s = 1.0$<br>$d_s = 0.1$ | No diffusion | $K = 50$<br>$q = 0$<br>$\mu = 0.001$ |
| A | $b_a = 1.1$<br>$d_a = 0.1$ | $D_a = 0.05$ | |
| I | $b_i = 0.3$<br>$d_i = 0.1$ | $D_i = 3.125$ | |

**Table 1 Parameters used in our simulations.** Concerning reaction velocities, unless specified, $r_{ab}(t) \equiv 1$ and $r_v(t)$ has the Hill form of equation (4) with $\eta = 3$.

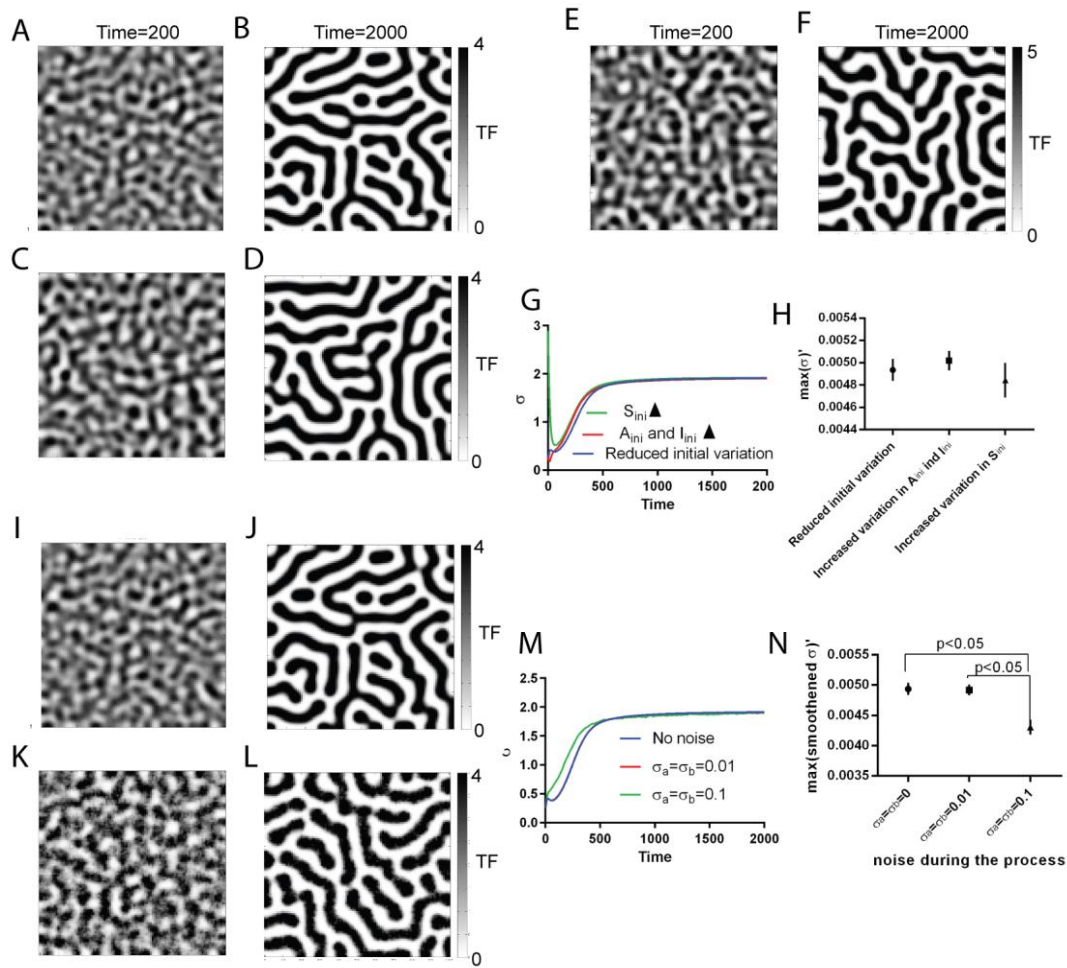

**Fig. 4 Patterns are not affected by initial and extrinsic noises.** This is illustrated here with parameters **A-B**) $A_{ini} = I_{ini} = [0.5, 0.6]$, $S_{ini} = [0.5, 1.5]$, and no noise during the process; **C-D**) $A_{ini} = I_{ini} = [0.5, 10.5]$, $S_{ini} = [0.5, 1.5]$, and no noise during the process; **E-F**) $A_{ini} = I_{ini} = [0.5, 0.6]$, $S_{ini} = [0.5, 10.5]$, and no noise during the process, **G**) Standard deviations of the three processes **H**) The steepness of $\sigma(t)$ curves is characterized by the maximum of $\sigma'(t)$ values. Data represent mean±standard deviation out of 5 independent simulation runs. One-way ANOVA tests show no significant differences between groups **I-L**) $A_{ini} = I_{ini} = [0.5, 0.6]$, $S_{ini} = [0.5, 1.5]$, and I-J) $\sigma_a = \sigma_i = 0.01$ K-L) $\sigma_a = \sigma_i = 0.1$ **M**) Standard deviations of the processes. **N**) The steepness of $\sigma(t)$ curves is characterized by the maximum of the first derivative of the smoothened $\sigma(t)$ values. Data represent mean±standard deviation out of 5 independent simulation runs. ANOVA + post hoc LSD test show significance. The images were taken at running times $t = 200$ (A,C,E,I,K) and $t = 2000$ (B,D,F,J,L). All other parameters are set according to Table 1 with a fixed domain size.

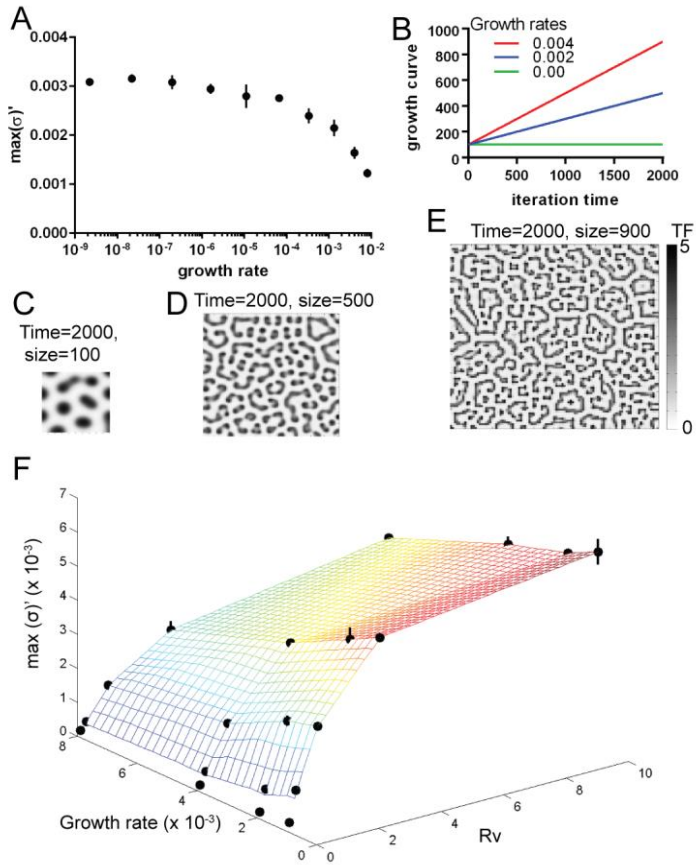

**Fig. 5. Effect of growth rate and reaction velocity on pattern formation**. The case of linear growth is investigated with a constant reaction velocity. Parameters are the ones of Table 1 with $b_a = 0.8$ and diffusions set to $D_a = 0.4$ and $D_i = 25$. The initial size of the domain is $M = 100$. A) Increased growth rates result in a decrease of the pattern formation. B) Linear growth were analyzed. C-E) Elevated growth rate also influence the pattern that are formed from regular dots to more and more lines. Larger growth rates promote the development of mixtures of lines and dots. F) Increased reaction velocity leads to higher maximal values of pattern development velocity independently of the growth rate, but the two factors mutually influence each other. A, F) Data represent mean±standard deviation out of 3 independent simulation runs.

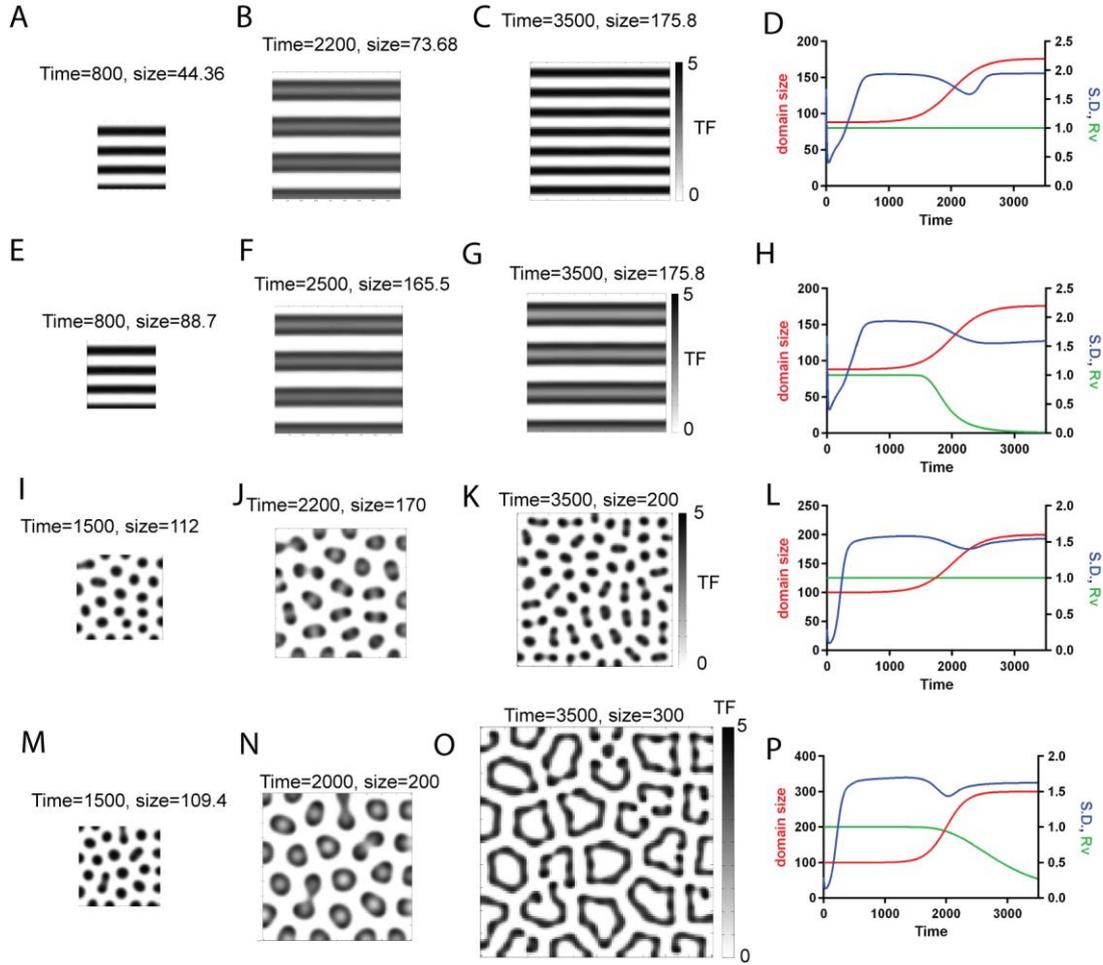

**Fig. 6 Illustrations of pattern formation on growing domains.** All parameters as in Table 1 unless specified; $D_a = 0.2$, and $D_i = 12.5$. Growth is of logistic form with $t_i = 2000$ and $r_g = 0.004$ and $\kappa = 2$ **A-D)** Zebra patterns develop from an initial domain size with $M = 88$ to $M = 176$ at time $t = 4000$. An initial linear gradient on [0.1,6] along the y-axis is used for $S(0, x)$, while $A_{ini} = I_{ini} = [0.5, 0.6]$. Domain growth is such that $t_i = 2000$ and $r_g = 0.004$, red trace in **D)** with a constant reaction velocity $r_v(t) \equiv 1$, green trace in **D)**. The standard deviation reacts continuously to growth and a new stabilization appear after each stripes divisions (blue trace). **E-H)** Same framework with a reaction velocity $r_v(t)$ of the Hill type, where $k_v = 500$ and $t_0 = 1400$. The standard deviation of the frozen pattern in **G)** at time $t = 3500$ remains smaller than the one of the fully developed pattern in C) at time $t = 3500$ or in E) at time $t = 800$. **I-L)** Formation of dots with $b_a = 0.7$. The reaction velocity is constant with time, $r_v(t) \equiv 1$. The initial conditions are uniformly random on $A_{ini} = I_{ini} = [0.5, 0.6]$ and $S_{ini} = [0.5, 1.5]$. Some dots are generated from already existing ones during growth periods as in **K)**, where $t = 3500$. **M-P)** Development of rosettes with $b_a = 0.8$. Growth is of logistic type with $t_i = 2000$ and $r_g = 0.006$ and $\kappa = 3$. Decreasing reaction velocity $r_v$ of the Hill type with $k_v = 1500$ and $t_0 = 1400$.